# APPROACH OF THE NGC 1977 STAR CLUSTER TO THE TOI-2796 HOST STAR

Hawi Yohanis Wakjira[1,2], D. A. Mosunova[3], E. S. Postnikova[3], N. V. Chupina[3],
S. V. Vereshchagin[3*]

[1]*Space Science and Geospatial Institute (SSGI), Entoto Observatory and Research Center (EORC),
Astronomy and Astrophysics Department, P. O. Box 33679 Addis Ababa, Ethiopia, hawiy@essti.gov.et*
[2]*Finfine Aerospace and Robotics International Institute (FARIS)
P.O. Box 12059, Addis Ababa, Ethiopia*
[3]*Institute of Astronomy of the Russian Academy of Sciences, Moscow, 119017 Russia*

*E-mail: svvs@ya.ru

## ABSTRACT

The subject of this article was the study of possible encounters in past epochs of the open star cluster NGC 1977 (from the group of clusters in Sword of Orion) with stars with planetary systems. For this purpose, the age of the cluster was determined based on our catalog data (Vereshchagin and Chupina 2023). Stars with planetary systems were selected from the NASA Archive catalog. The age of the cluster was determined using the color–absolute magnitude diagram and the isochron system. By extending the track of the movement of the cluster and stars in past epochs (≈10 Myr), the time of the maximum approach (≈32 pc) of the host star with planetary system TOI-2796 with the NGC 1977 are found. The place of approach in the sky is shown - this point can be considered as the place of appearance of interstellar comets. Thus, the result of our work is that the we found approach of the host star to the cluster entailed effects associated with the gravitational influence of the cluster on the nuclei of comets located in the outer parts of the Oort cloud of the planetary system. The effect of approach on comets in the Oort cloud is estimated.

## INTRODUCTION

Close passages of stars in past epochs could have affected the orbital elements of cometary nuclei in the Oort cloud. The gravitational effect of such encounters leads to both the appearance of new comets in the Solar System and the appearance of interstellar asteroids (Torres et al. 2019, Vereshchagin et al. 2022). The theory of this effect was developed by (Rickman 1976). Not only stars and the Sun, but also star clusters and the Sun could undergo rapprochements in space. Thus, Vereshchagin et al. (2022) discovered the approach of the Sun to the Hyades about two million years ago. The task of searching for encounters between star clusters and other planetary systems in the Galaxy is of undoubted interest.

The Gaia DR3 star data allows the selection of stars with the most reliable phase space coordinates. Our interest is the young (age ≈ 15 Myr) open star cluster (OSC) NGC 1977 (UBC 621, MWSC 0587). That cluster is located in the Orion Sword region at a distance of about 400 pc from the Sun. We used the catalog of candidate stars for a cluster (Vereshchagin and Chupina 2023, VC2023), that makes it possible to determine the main parameters necessary for calculating its motion in space: spatial velocity, age, and its size. These data make it possible to calculate the position of the cluster in past epochs and the parameters of its approach to surrounding stars. Obviously, stars with planetary systems, which are contained in the Exoplanet NASA Archive (URL: http://exoplanetarchive.ipac.caltech.edu), are of particular interest. For this purpose, stars were selected that have planetary systems and could potentially approach the cluster in past epochs.

All this suggests that in the Orion Sword region we have a natural laboratory that allows us to investigate both the origin of practically unexplored multiple clusters and exotic stars, in particular, with planetary systems.

Work structure. The Data section contains the selection of stars from the catalog VC2023 with a probability of belonging greater than 60%, and an assessment of the quality of the data of individual stars. Next, in order to most reliably estimate the position of the center of the cluster and its size, the distribution of parallaxes and the distribution of stars in space are considered. Next, it is estimated the age of the cluster using isochrones. In a separate section, stars with planetary systems located in the vicinity of the cluster are considered. The size of the neighborhood is taken to be sufficient, taking into account the movement of stars and clusters in past epochs and the search for their possible convergence in space. In the final part, the selected objects were moved in space in past epochs and the parent star that was closest to the cluster was found. In conclusion, the effects of the approach were assessed and a discussion was carried out and conclusions were drawn to conclude that the cluster could have an impact on the planetary system of the star TOI-2796.

## DATA

VC2023 catalog, used in that work, contain membership stars both in NGC 1977 and in NGC 1981 (n=1041) with corresponding probabilities. The clusters in space almost completely overlap with each other. The centers of the OSCs NGC 1981 and NGC 1977 are located within a region of the sky ~0.5° (~5 pc) in size. The Fig.1 shows the distribution of stars in the equatorial coordinate system.

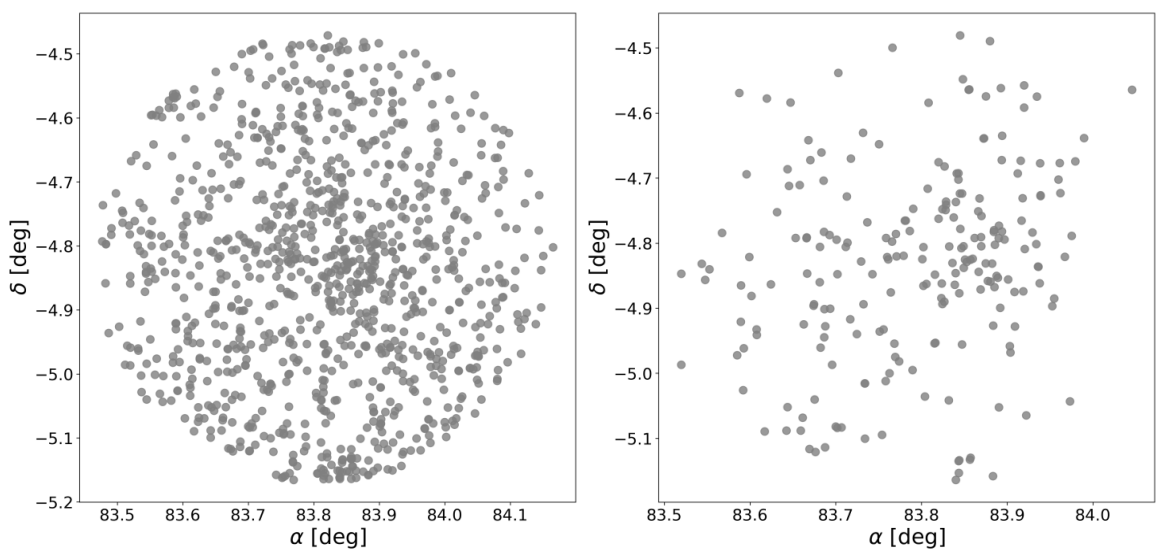

Fig.1. The distribution of the NGC 1977 stars in the equatorial coordinate system: all stars (n=1041) – left panel and stars with membership probability P_NGC1977≥60% (n=217) – right panel.

Our datasets include not only probable members with P ≥60%, but with most reliable parallax measurements (Fig. 2). Fig.2 clearly shows the possible parallax uncertainties inherited Gaia DR3 from Gaia DR2. It is likely that the use of the maximum time interval has not yet been reached and the PSF modeling used in image location has not been updated, Gaia collaboration, Brown, A. G. A. et al. 2018.

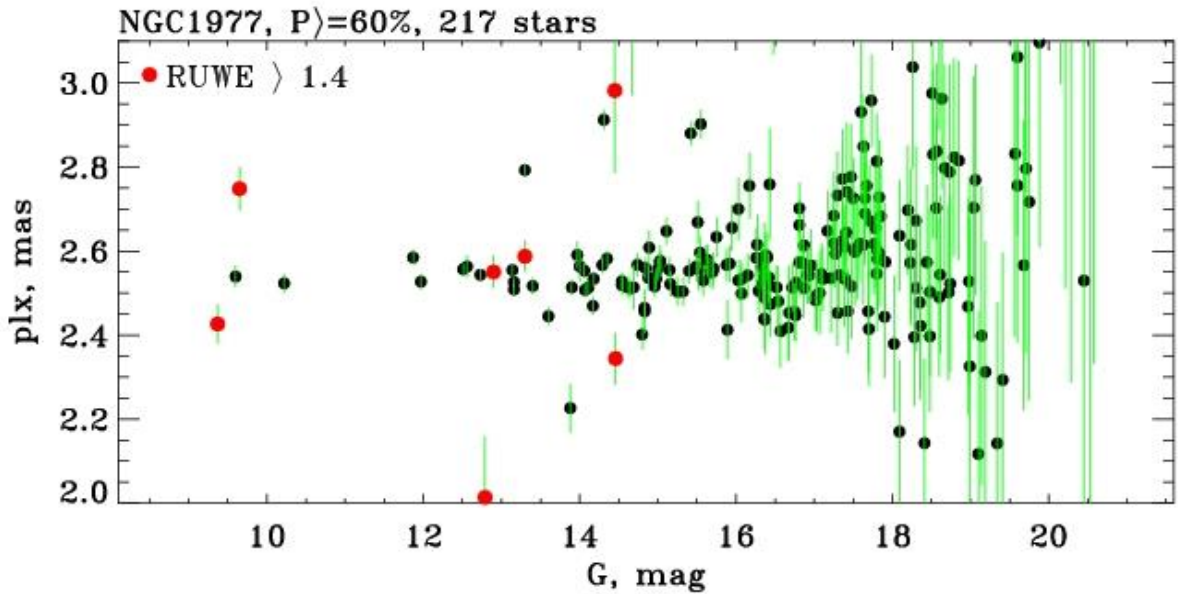

Fig.2. It is shown how the scatter of the parallax values (with error bars shown in green) of cluster membership candidates depends on G magnitude Gaia DR3. Red points – stars with RUWE>1.4 (n=9) and G<16.

In Fig. 2 we see 4 stars brighter than G=14 (n=22) whose parallaxes deviate the most from the mean. What's the matter? Obviously, there are some percent of background stars inside the region of space occupied by the cluster. Such stars may be absolutely no difference from cluster membership stars. In particular, to have proper motion and parallax values indistinguishable from cluster stars, which are selected just according to these parameters. In VC2023 it was estimated the fraction of such stars. The 4 stars in question account for 18% of the sample of 22 stars, which fits well into the estimates given in (VC2023, Fig.10). We assume that these 4 stars with bouncing parallaxes in Fig. 2 belong to the background and do not belong to the cluster.

We made 3 samples:

**dataset1** includes stars with $P_{NGC1977}$ ≥60%, n=217, they located at a distance from the Sun from 273 and to 519 pc.

**dataset2** includes stars with $P_{NGC1977}$ ≥60%, G<14 and 2.3<π<2.7 mas, n=18. In this sample we also used stars with RUWE <1.4, marked in Fig. 2. Based on this sample, we determined the average space velocity, the center of the cluster and, most importantly, the radius of the region of space in which its stars are located.

**dataset3** includes stars with $P_{NGC1977}$ ≥60% and G<16, n=74.

## NGC 1977 POSITION IN THE GALAXY, SPATIAL VELOCITY COMPONENTS AND CLUSTER CENTER

The spatial coordinates and velocity components: X, Y, Z (and U, V, W) match the Galactic heliocentric Cartesian reference frame with X-axis directed to the Galactic center ($l = 0°, b = 0°$), Y-axis, to the Galaxy rotation direction ($l = 90°, b = 0°$), and Z-axis, to the Northern Galactic Pole ($l = 90°, b = 90°$).

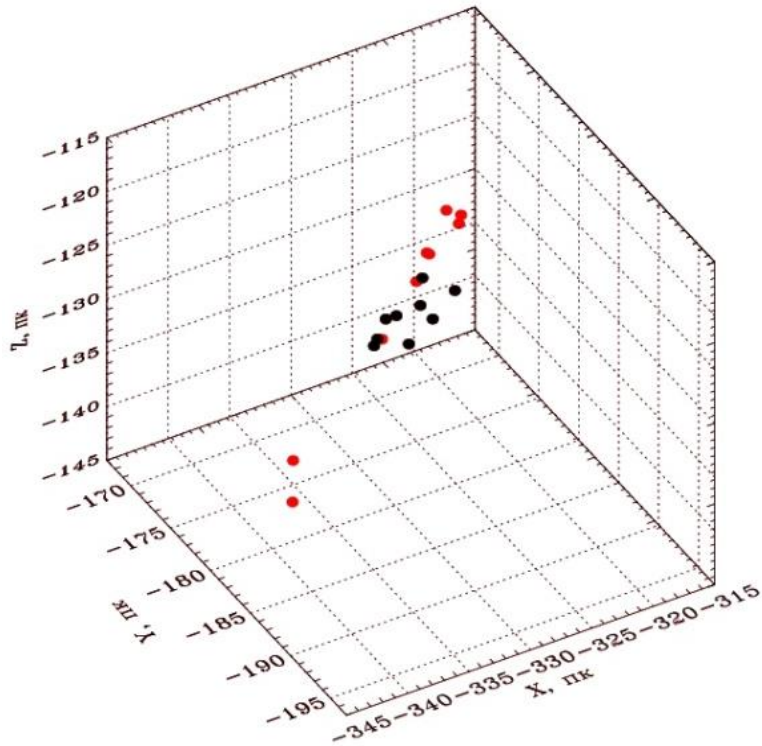

Fig.3. 3D distribution of dataset2 stars. Stars with RUWE>1.4 are shown with red dots.

For stars of dataset2, we took the parameters from Gaia DR3, see Table 1. Using these data, we determined spatial coordinates and velocity components. These data are presented in the Table 2.

Columns of Table 1 contain: ID Gaia DR3, probability of a star belonging to a cluster $P_{NGC1977}$, right ascension α (J2000), declination δ (J2000), renormalised unit weight error RUWE, parallax π, parallax error $\varepsilon_\pi$, proper motion $\mu_\alpha$, proper motion error $\varepsilon_{\mu\alpha}$, proper motion $\mu_\delta$, proper motion error $\varepsilon_{\mu\delta}$, color index BP-RP, G magnitude, error of G, extinction in G band AG, reddening E(BP-RP), iron abundance [Fe/H], Dist – distance from the Sun using BP/RP spectra, radial velocity Vr, error of Vr.

Table 1. Dataset2 data from Gaia DR3 (n=18) and 4 stars having parallax beyond the interval 2.3<π<2.7 mas (marked with '*'). Two stars that are on Fig. 3 is far from all the others are marked with '**'.

| GDR3 | P% | α,° | δ,° | RUWE | π | $\varepsilon_\pi$ | $\mu_\alpha$ | $\varepsilon_{\mu\alpha}$ | $\mu_\delta$ | $\varepsilon_{\mu\delta}$ | BP-RP | G | $\varepsilon$G mag | AG mag | E(BP-RP) mag | A0 mag | [Fe/H] | Dist pc | Vr km/s | $\varepsilon$Vr km/s |
|---|---|---|---|---|---|---|---|---|---|---|---|---|---|---|---|---|---|---|---|---|
| 3209526874441202560 | 65 | 83.67617 | -5.12072 | 1.140 | 2.5395 | 0.0260 | 1.197 | 0.021 | -0.150 | 0.018 | 0.041901 | 9.60 | 0.002770 | 0.1370 | 0.0729 | 0.1425 | -1.4937 | 423.5919 | 23.30 | 2.35 |
| 3209528081326372736* | 94 | 83.88320 | -5.15785 | 3.956 | 2.0136 | 0.1446 | 1.137 | 0.242 | -1.005 | 0.227 | 1.771276 | 12.79 | 0.003426 | | | | | | | | |
| 3209528493645505920 | 72 | 83.84356 | -5.15308 | 1.053 | 2.5549 | 0.0126 | 1.011 | 0.012 | -1.299 | 0.010 | 1.438042 | 13.14 | 0.002813 | | | | | | | 20.81 | 1.86 |
| 3209545085102536192 | 60 | 83.64362 | -5.05192 | 1.249 | 2.5233 | 0.0228 | 1.308 | 0.019 | -1.503 | 0.016 | 0.371922 | 10.22 | 0.002786 | 0.2602 | 0.1392 | 0.2840 | -0.5586 | 415.4450 | 33.89 | 1.63 |
| 3209549418726081536 | 66 | 83.59270 | -4.96140 | 1.000 | 2.5435 | 0.0134 | 1.101 | 0.011 | -1.420 | 0.009 | 1.269306 | 12.73 | 0.003179 | 0.4914 | 0.2641 | 0.6208 | -0.4107 | 379.5829 | 26.54 | 5.34 |
| 3209571958715287680* | 67 | 83.90096 | -4.85582 | 1.725 | 2.2263 | 0.0578 | 1.902 | 0.053 | -0.736 | 0.044 | 2.078219 | 13.88 | 0.003602 | 2.5914 | 1.4462 | 3.3938 | -1.1896 | 456.5980 | 23.91 | 7.75 |
| 3209572405390979840** | 65 | 83.80561 | -4.86248 | 2.653 | 2.4265 | 0.0466 | 1.421 | 0.039 | -1.437 | 0.035 | 0.733916 | 9.37 | 0.002929 | | | | | | | | |
| 3209572852067565568 | 91 | 83.83964 | -4.84591 | 1.405 | 2.5849 | 0.0187 | 1.079 | 0.016 | -1.026 | 0.015 | 1.482145 | 11.87 | 0.005715 | 1.8719 | 0.4671 | 1.1253 | -0.1399 | 382.1849 | 22.61 | 4.76 |
| 3209574776212896384 | 83 | 83.86203 | -4.79188 | 1.284 | 2.5273 | 0.0181 | 1.463 | 0.015 | -1.180 | 0.013 | 1.341905 | 11.97 | 0.003835 | 0.7060 | 0.3804 | 0.8918 | -0.3383 | 378.6376 | 19.90 | 4.81 |
| 3209575703925864576 | 74 | 83.76110 | -4.82483 | 1.777 | 2.5908 | 0.0321 | 1.377 | 0.026 | -1.334 | 0.023 | 2.427351 | 13.97 | 0.003979 | | | | | | | 24.35 | 5.60 |
| 3209576597279031680 | 78 | 83.81973 | -4.76239 | 1.991 | 2.5638 | 0.0368 | 0.808 | 0.030 | -0.659 | 0.025 | 1.817097 | 14.00 | 0.026292 | | | | | | | 26.53 | 3.78 |
| 3209576769077759744 | 67 | 83.71368 | -4.79919 | 2.463 | 2.5503 | 0.0378 | 0.653 | 0.035 | -0.829 | 0.029 | 1.712741 | 12.90 | 0.003205 | | | | | | | | |
| 3209576769077761152 | 65 | 83.70098 | -4.79449 | 1.665 | 2.5169 | 0.0205 | 0.744 | 0.019 | -0.425 | 0.015 | 2.195463 | 13.40 | 0.003266 | | | | | | | | |
| 3209577937308816896 | 78 | 83.86609 | -4.75094 | 1.055 | 2.5087 | 0.0136 | 1.768 | 0.012 | -0.752 | 0.011 | 1.525757 | 13.16 | 0.002906 | | | | | | | 29.96 | 32.70 |
| 3209578074747777792 | 96 | 83.82441 | -4.74793 | 1.009 | 2.5136 | 0.0183 | 1.255 | 0.015 | -0.605 | 0.012 | 1.746463 | 13.90 | 0.003929 | 1.5737 | 0.8666 | 2.0458 | -1.0290 | 385.9929 | 198.96 | 13.45 |
| 3209578276615162112 | 92 | 83.84715 | -4.72385 | 1.868 | 2.5633 | 0.0260 | 1.180 | 0.021 | -1.066 | 0.017 | 1.514975 | 12.56 | 0.003013 | 1.2407 | 0.6746 | 1.5779 | -0.5973 | 411.8603 | 22.15 | 5.19 |
| 3209579724015206272 | 62 | 83.81989 | -4.67570 | 2.709 | 2.5176 | 0.0366 | 0.993 | 0.033 | -0.176 | 0.028 | 1.988200 | 13.30 | 0.003448 | | | | | | | | |
| 3209579772401520628* | 99 | 83.82685 | -4.68219 | 3.033 | 2.7485 | 0.0535 | 1.338 | 0.048 | -0.734 | 0.038 | 0.481926 | 9.65 | 0.002764 | | | | | | | | |
| 3209598587511657216 | 77 | 83.55263 | -4.84019 | 1.030 | 2.5569 | 0.0116 | 1.315 | 0.011 | -0.291 | 0.009 | 1.439836 | 12.51 | 0.003216 | 0.8344 | 0.4466 | 1.0694 | -0.1004 | 388.5616 | 25.51 | 38.16 |
| 3209624529114069888 | 78 | 83.71278 | -4.72818 | 1.005 | 2.5127 | 0.0127 | 1.766 | 0.011 | -0.718 | 0.010 | 1.624669 | 13.16 | 0.002892 | | | | | | | | |
| 3209624872711453696* | 81 | 83.68561 | -4.70405 | 1.018 | 2.7929 | 0.0127 | 1.061 | 0.011 | -1.195 | 0.010 | 1.373308 | 13.30 | 0.003052 | 0.4086 | 0.2188 | 0.5275 | -0.4610 | 353.9284 | 28.20 | 3.21 |
| 3209625044510130304** | 60 | 83.71755 | -4.66994 | 1.694 | 2.4449 | 0.0213 | 1.145 | 0.020 | -1.494 | 0.016 | 1.782487 | 13.60 | 0.013776 | | | | | | | 20.45 | 2.60 |

Columns of Table 2 contain: ID Gaia DR3, probability of a star belonging to a cluster $P_{NGC1977}$, renormalised unit weight error RUWE, spatial coordinates x,y,z, velocity components U,V,W.

Table 2. Parameters we calculated for the dataset2 stars

| GDR3 | P% | RUWE | x pc | y pc | z pc | U km/s | V km/s | W km/s |
|---|---|---|---|---|---|---|---|---|
| 3209526874441202560 | 65 | 1.140 | -325.8 | -178.3 | -130.7 | 19.22 | -11.96 | -5.98 |
| 3209528493645505920 | 72 | 1.053 | -323.8 | -178.0 | -129.1 | 15.94 | -12.18 | -6.32 |
| 3209545085102536192 | 60 | 1.249 | -328.2 | -179.0 | -131.5 | 26.59 | -18.64 | -10.39 |
| 3209549418726081536 | 66 | 1.000 | -325.9 | -176.9 | -130.5 | 20.61 | -14.93 | -8.22 |
| 3209572405390979840 | 65 | 2.653 | -342.1 | -185.8 | -135.2 | 0 | 0 | 0 |
| 3209572852067565568 | 91 | 1.405 | -321.2 | -174.5 | -126.7 | 17.80 | -12.60 | -6.55 |
| 3209574776212896384 | 83 | 1.284 | -328.7 | -178.2 | -129.3 | 15.41 | -12.02 | -5.15 |
| 3209575703925864576 | 74 | 1.777 | -320.5 | -173.7 | -126.8 | 18.97 | -14.05 | -6.93 |
| 3209576597279031680 | 78 | 1.991 | -324.2 | -175.4 | -127.6 | 21.43 | -13.60 | -7.95 |
| 3209576769077759744 | 67 | 2.463 | -325.7 | -176.1 | -129.0 | 0 | 0 | 0 |
| 3209576769077761152 | 65 | 1.665 | -330.0 | -178.4 | -130.8 | 0 | 0 | 0 |
| 3209577937308816896 | 78 | 1.055 | -331.3 | -179.4 | -130.1 | 24.25 | -16.29 | -7.58 |
| 3209578074747777792 | 96 | 1.009 | -330.7 | -178.9 | -130.1 | 164.84 | -91.54 | -63.56 |
| 3209578276615162112 | 92 | 1.868 | -324.4 | -175.4 | -127.4 | 17.41 | -12.54 | -6.25 |
| 3209579724015206272 | 62 | 2.709 | -321.5 | -173.4 | -126.2 | 0 | 0 | 0 |
| 3209598587511657216 | 77 | 1.030 | -324.7 | -175.3 | -129.7 | 20.98 | -13.14 | -6.65 |
| 3209624529114069888 | 78 | 1.005 | -329.1 | -177.5 | -130.0 | 0 | 0 | 0 |
| 3209625044510130304 | 60 | 1.694 | -340.3 | -183.1 | -134.2 | 15.50 | -12.42 | -6.09 |

To determine cluster center space coordinates and their dispersion ($X_0 \pm \sigma_X, Y_0 \pm \sigma_Y, Z_0 \pm \sigma_Z$) we averaged the coordinates of individual stars. Table 3 shows the results of these calculations.

Table 3. Spatial coordinates of the cluster center of the on various lists.

|  | dataset2, n=18 | | dataset2, RUWE<1.4, n=9 | |
|---|---|---|---|---|
|  | mean | dispersion | mean | dispersion |
| X | -327.67 | 5.88 | -327.58 | 2.65 |
| Y | -177.63 | 3.13 | -177.94 | 1.26 |
| Z | -129.72 | 2.40 | -130.11 | 0.73 |

For stars dataset2 having measurements in Gaia DR3 the RV (n=13) we have calculated the average values of spatial velocity component (U±σ_U,V±σ_V,W±σ_W), Table 4. For calculations we also used dataset2 and RUWE <1.4. Table 4 contains the obtained values of the spatial velocity components of the cluster and their dispersions. The star GDR3 3209578074747777792 was not taken into account in the calculations, since it has an inexplicably high radial velocity (Table 1).

Table 4. Average space velocity and dispersion of membership stars NGC 1977, dataset2.

|  | dataset2, n=12 | | dataset2, RUWE<1.4, n=7 | |
|---|---|---|---|---|
|  | mean | dispersion | mean | dispersion |
| U | 19.51 | 3.48 | 20.43 | 4.08 |
| V | -13.70 | 2.03 | -14.16 | 2.57 |
| W | -7.00 | 1.38 | -7.18 | 1.74 |

The NGC 1977 spatial velocity is equal $V_{NGC1977} \approx 24.8$ km/s, from Table 4 data (dataset 2).

**COLOR-MAGNITUDE DIAGRAM, AGE ESTIMATION**

Fig. 4 shows color-magnitude diagrams (CMD) constructed for candidates for the cluster members (dataset1) and the most probable members (dataset2). The diagrams were constructed without taking into account the absorption of light by the interstellar medium.

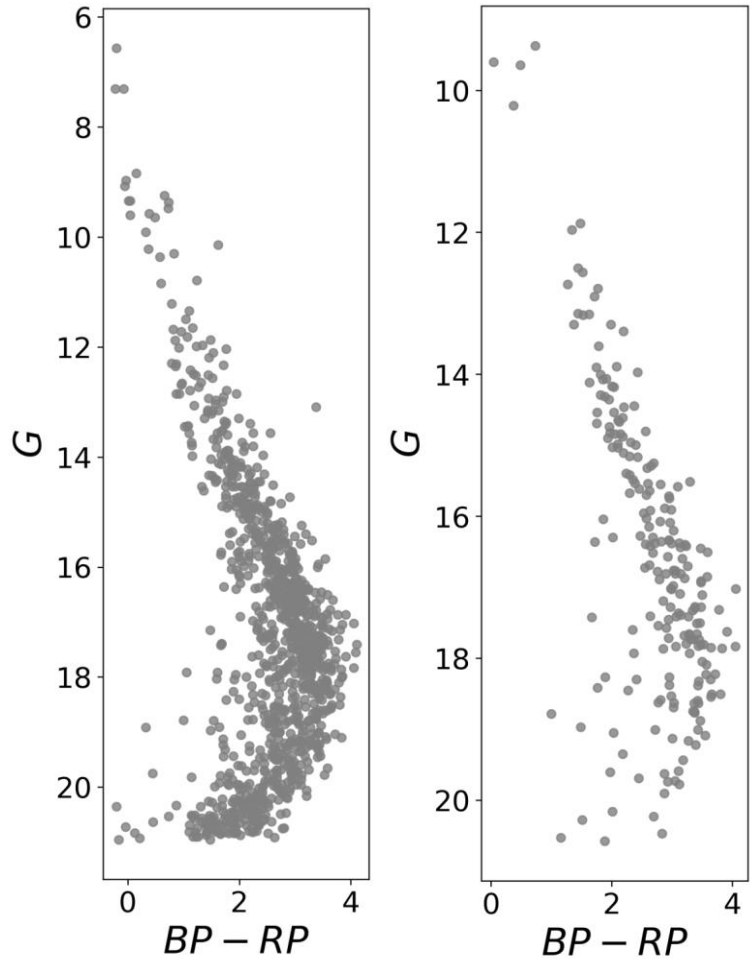

Fig.4. Color-magnitude diagram (n = 1041) for NGC 1977 – left panel and dataset1 (stars with $P_{NGC1977}$ ≥60% membership probability, n=217) – right panel.

In order to solve the problem of determining the age of the cluster, it is necessary to take into account the absorption of light by the interstellar medium and obtain non-reddened color indices of stars and determine their absolute magnitudes. We take from Gaia collaboration, Gaia DR3 (I/355 catalog) "extinction in G band from GSP" - AG, "reddening E(GBP-GRP) from GSP" - E(BP-RP) and calculate the absolute magnitude and extinction. This is how the color index is corrected:

$M_G$=Gmag+5+5(lg π) − $A_G$

(BP-RP)$_0$=(BP-RP)-E(BP-RP)

On Fig.5 we construct the three diagrams $M_G$-$(BP - RP)_0$. The number of stars in the samples has decreased significantly. This is due to the presence of absorption coefficients not for all stars.

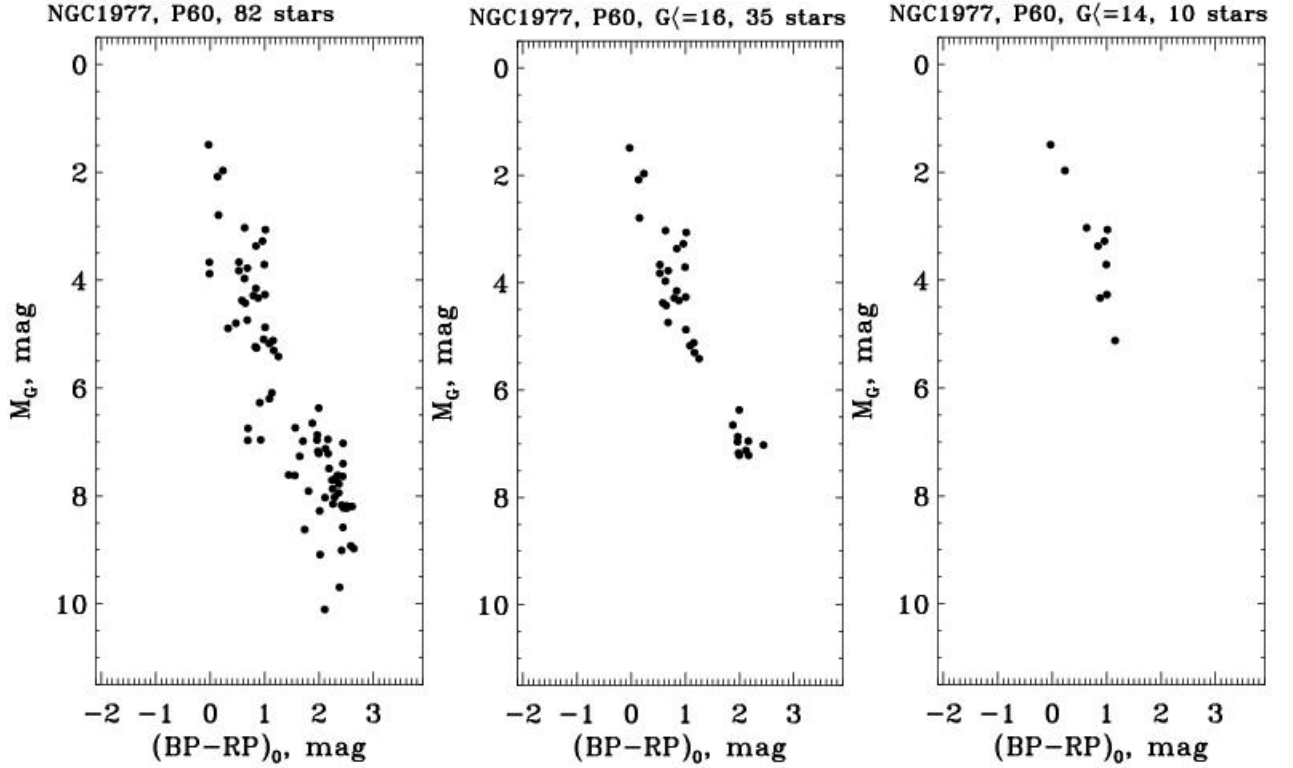

Fig.5. The $M_G - (BP - RP)_0$ for dataset1 – left (82 stars from 217 have AG and E(BP-RP)), for dataset3 – on the middle (n=35), for dataset2 – on the right (n=10).

The cluster age according to the isochron. The $M_G$-$(BP - RP)_0$ diagram on Fig. 5 allows us to put all stars of a given cluster in a CMD and to determine their age by comparing with stellar isochrones. For this purpose, we acquired a grid of solar-metallicity PARSEC isochrones (version 1.2S). Isochrones on Fig. 6 are taken from the Padua server: http://stev.oapd.inaf.it/cgi-bin/cmd. It is selected the Photometric system Gaia EDR3 (all Vegamags, Gaia passbands from ESA/Gaia website) for ages = $1.2 \cdot 10^6$ - $16.8 \cdot 10^6$ yr. We consider that G in Gaia that is $M_G$ and $(BP - RP)_0$ is $G_{BP} - G_{RP}$ respectively in Padova designation.

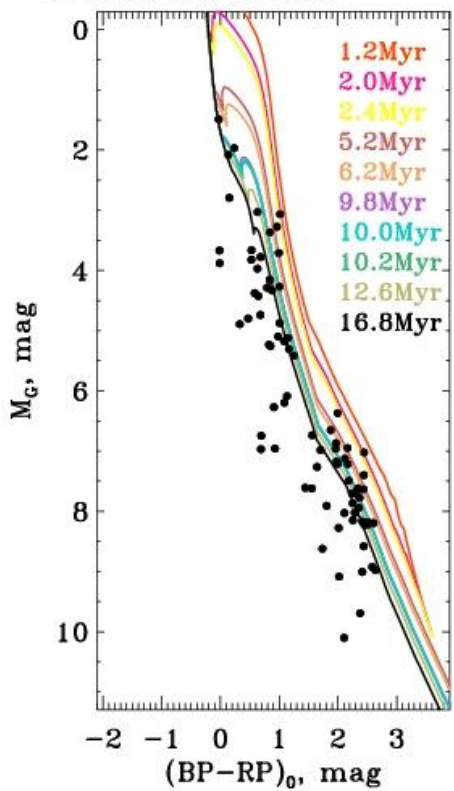

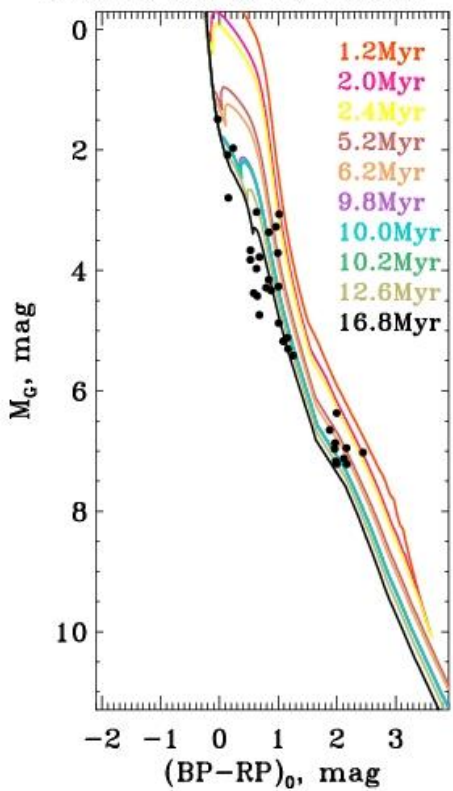

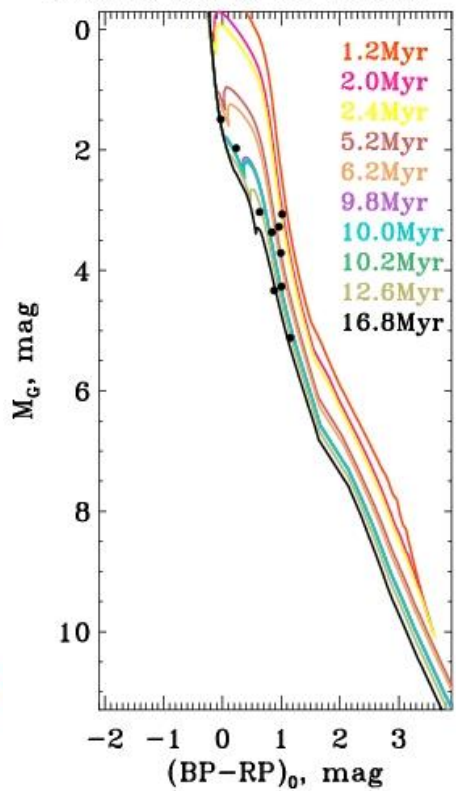

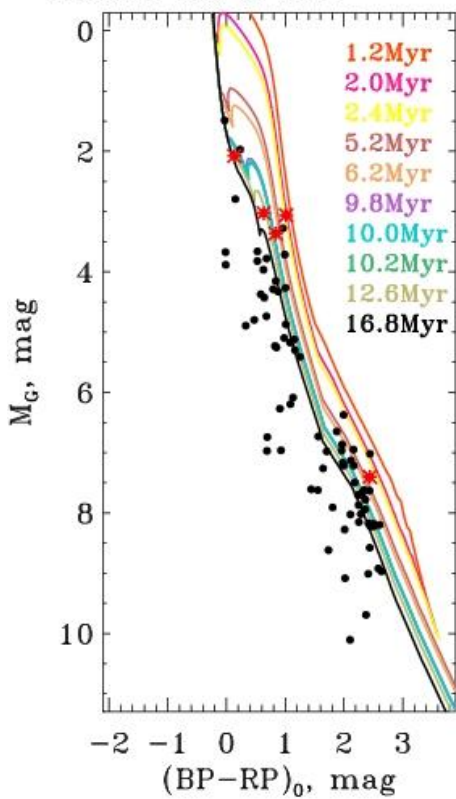

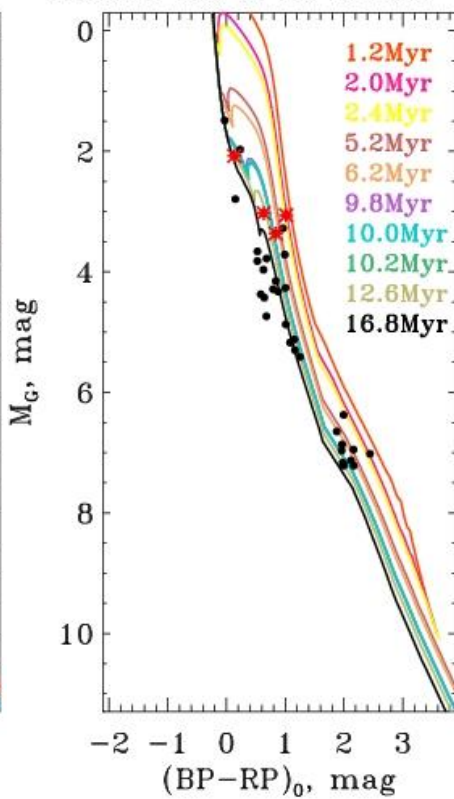

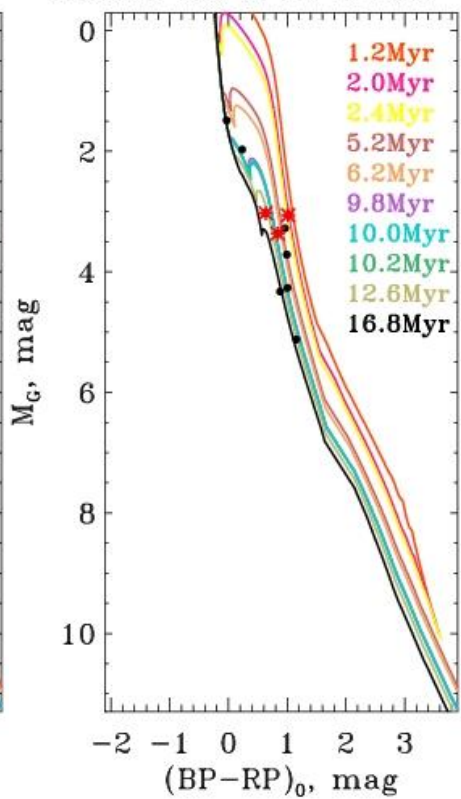

Fig.6. The panels with an isochrone grid placed on each panel. The color–absolute magnitude diagram showing all the identified members of the NGC 1977 in black. Stars with RUWE>1.4 shown by red dots. A grid of solar-metallicity PARSEC isochrones (version 1.2S). It is got the Table of values and table "Useful data": ages = 1.0e7, 1.0e6 and 1.0e8. Number of stars with RUWE>1.4: 5 from n=82 (left); 4 from n=35 (in the middle); 3 from n=10 (right).

The age of NGC 1977 according to Cantat-Gaudin et al. (2020) is 7.99 (97.7 Myr). Another estimate is 3 Myr Pang et al. (2022). According to MWSC Kharchenko et al. (2013), the age of the NGC 1977 cluster is 6.600 (4 Myr). A close estimate with 6.721 ± 0.064 (5.3 Myr) was obtained (Monteiro, et al. 2020). Obviously, the above estimates do not allow us to speak about the similarity of these OSC in terms of age. The nearest isochrone in the middle panel of the Fig.6 is plotted for an age of 10 million years, which is our age estimate for NGC 1977.

## STARS WITH PLANETARY SYSTEMS IN THE VICINITY OF THE CLUSTER

Data on host stars was taken from the Exoplanet NASA Archive located on URL= http://exoplanetarchive.ipac.caltech.edu. It has its own ability to select stars according to the parameters indicated in the column names. We have searched the unique host stars located in alpha from 70 to 100 and in delta from -15 to 5 degrees (approximately 100 pc from the cluster center). We found three stars with planetary systems, Table 5. In columns of the Table 5 are the name of star with planetary system – hostname, equatorial coordinates – $\alpha$ and $\delta$, distance from the Sun – sy_dist and distance error interval limits – sy_disten1 and sy_disten2.

Table 5. Stars with planetary systems (host stars) located within about 100 pc from the center of the cluster NGC 1977.

| hostname | gaia_id | $\alpha$ | $\delta$ | sy_dist | sy_disterr1 | sy_disterr2 |
|----------|---------|----------|----------|---------|-------------|-------------|
| Gaia-1 | Gaia-1 | 90.6436658 | -0.5771099 | 363.660 | 3.058 | -3.008 |
| TOI-2796 | Gaia DR2 3222453935725951488 | 84.1527191 | 0.8962788 | 350.343 | 6.018 | -5.821 |
| TOI-892 | Gaia DR2 3010759262610139776 | 86.738228 | -11.2353401 | 340.543 | 4.894 | -4.76 |

## APPROACH THE HOST STAR TOI-2796 AND NGC1977

The tracks of star and clusters space motion in past epochs are considered. Linear approximation is used, Fig. 7. Data on spatial coordinates (x y z) and spatial velocity components (U,V,W) given in Table 6. The $\mu_\alpha$ and $\mu_\delta$, distance from the Sun, radial velocity (Vr) are taken from SIMBAD for the NGC 1977 cluster and from the Archive NASA for the stars.

Table 6. Initial data the tracks of space motion of stars and cluster.

| hostname | GAIA_ID | α | δ | l | b | sy_dist | x | y | z | μα | μδ | Vr | U | V | W |
|---|---|---|---|---|---|---|---|---|---|---|---|---|---|---|---|
| Gaia-1 | Gaia-1 | 90.6436658 | -0.5771099 | 207.718 | 3.6254 | 363.660 | -321.28488 | -168.80714 | 22.995295 | 5.989 | -40.632 | | | | |
| TOI-2796 | Gaia DR2 3222453935725951488 | 84.1527191 | 0.8962788 | 203.63 | 3.554 | 350.343 | -320.35049 | -140.15750 | 21.717491 | 2.537 | 2.635 | 20.91 | 19.52 | -6.55 | 7.08 |
| TOI-892 | Gaia DR2 3010759262610139776 | 86.738228 | -11.2353401 | 216.165 | 3.7728 | 340.543 | -274.33170 | -200.52319 | 22.407803 | -0.282 | 6.035 | 41.10 | 38.23 | -16.78 | 6.40 |
| NGC 1977 | | 83.8150 | -4.8190 | | | 396.67 | -329.45 | -178.65 | -129.98 | 1.271 | -0.753 | 24.20 | 19.41 | -13.19 | -6.53 |

In Fig.7 (and Table 7) we see that the NGC 1977 cluster approached the Gaia DR2 3222453935725951488 (TOI-2796) about a ten million years ago on the distance ~33 pc. During the approach, there was an influence on the planet system, especially on comets in its outer regions. This event change the major axis of comets' orbits and result in their subsequent close approach star and star cluster or ejection to interstellar medium.

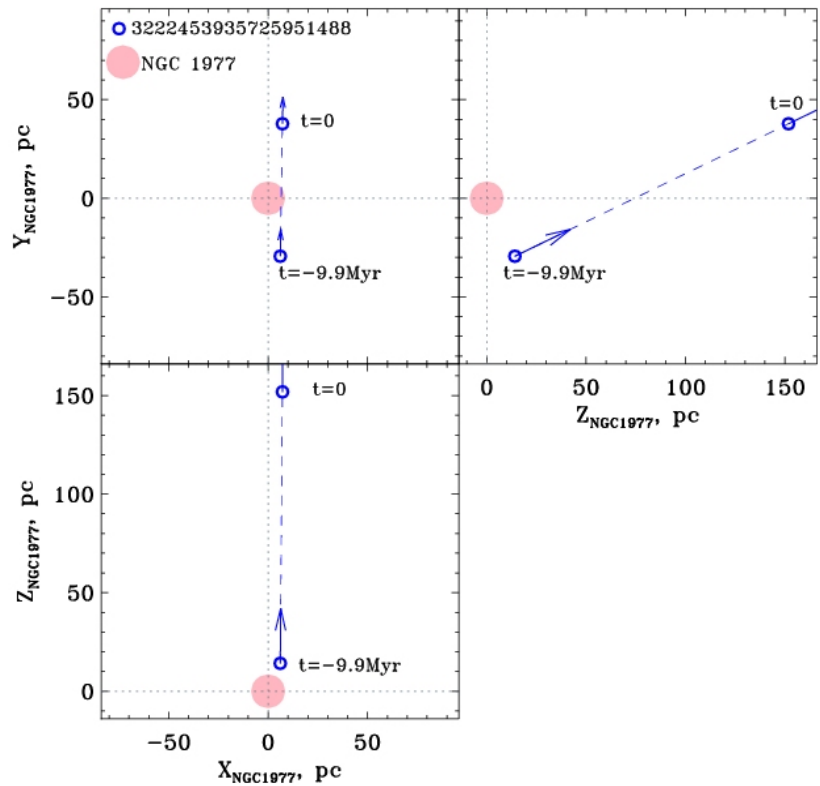

Fig.7. The star TOI-2796 on XYZ and a cluster move in early epoch (from t=0 to t=9.9 Myr). The cluster is taken at the center of the rectangular coordinate system.

Table 7. Resulting approach parameters: $t_{min}$ time and distance $d_{min}$ of approach.

| ID | TESS | t_min, yr | d_min, pc |
|---|---|---|---|
| Gaia DR2 3222453935725951488 | TOI-2796 | 9.90e+6 | 32.9 |
| Gaia DR2 3010759262610139776 | TOI-892 | 5.65e+6 | 94.4 |

For us, the star of TOI-2796, which came closer to the NCC 1977 at a distance of 32.9 pc (Table 7), is of interest. Planet TOI-2796 b has Planetary Parameters in NASA Exoplanet Archive (URL=https://exoplanetarchive.ipac.caltech.edu/overview/TOI-2796). The semi-major axis of the TOI-2796 b (Hot Jupiter) orbit around star TOI-2796 is equal 0.0569+0.0010-0.0011. The mass of this star is almost equal one solar mass, Exoplanet NASA Archive. The latter allows to evaluate the radius of clouds of cometary nuclei, quite possibly filling the outer part of TOI-2796 planetary system. By analogy with the solar system, we will call it a cloud of Oort. His radius, which will be needed below, we appreciated in the usual way. The $R_{Oort}$ – the radius of TOI-2796 Oort cloud. Determine the radius of the Oort cloud or dust shell (which was here 10 million years ago) for TOI-2796. The Hill radius or sphere has been described as "the region around a planetary body where its own gravity is the dominant force of attraction for the satellites". The outer boundary of the Oort cloud (Oort 1950) determines the gravitational boundary of the planet system - the Hill sphere, (defined for the solar system at 0,6 pc, or 120 AU). The Hill radius, Lissauer and Murray (2014):

$$R_{Hill} = a \cdot \sqrt[3]{\frac{m}{3(M_{Gal}+m)}} \ , \ \ \text{pc} \qquad (1)$$

were a – semi-major axis of the star orbit around the galactic Center. The distance from the Sun to the Galactic Center is equal $R_0$=8.178±0.013(stat) ± 0.022(syst) kpc, The GRAVITY collaboration (2019), $M_{Gal}$ – mass of the Galaxy, $M_{Gal}$=1.15·$10^{12}M_{\odot}$ (Watkins et al. 2019). The star TOI-2796 mass m=1.063 (+0.057-0.062), age is equal 4.0 (+3.3-2.5), Yee et al. (2023). Mass m=0.937 – 1.017, mean=0.977, age (Gyr) from 5.427 to 7.641, mean 6.542, SIMBAD (Gaia collaboration , I/355/paramp). Substituting these values into the (1), we get $R_{Hill} = R_{Oort}$=0.55 pc.

## EFFECTS OF SPATIAL APPROACH

Let $D_0$ is the distance from the star to the comet, from which it can be ejected into interstellar space, Mülläri & Orlov (1996), Bobylev (2017). Below this distance star's gravitational attraction prevails. Close to $D_0$, the gravitational forces acting on the comet from an approaching object begin to exceed the gravity of the star. $D_0$ is equal to:

$$D_0 = d_{min}(1 - \frac{1}{1+\sqrt{1/M_{cluster}}}) \qquad (2)$$

where $M_{cluster}$ is the mass of the passing object (is it an individual star or a cluster as a whole, in $M_{\odot}$).

Fig. 8 shows, how $d_{min}$ affects the dependence of $D_0$ on the mass $M_{cluster}$ (2) of the approaching object. In case $D_0 \lesssim R_{Oort}$ the comet does leave the Oort cloud. In Fig. 8 the picture of dependence is presented for various values of the mass of star cluster and the minimum distance, at which the cluster center and the star are as close as possible.

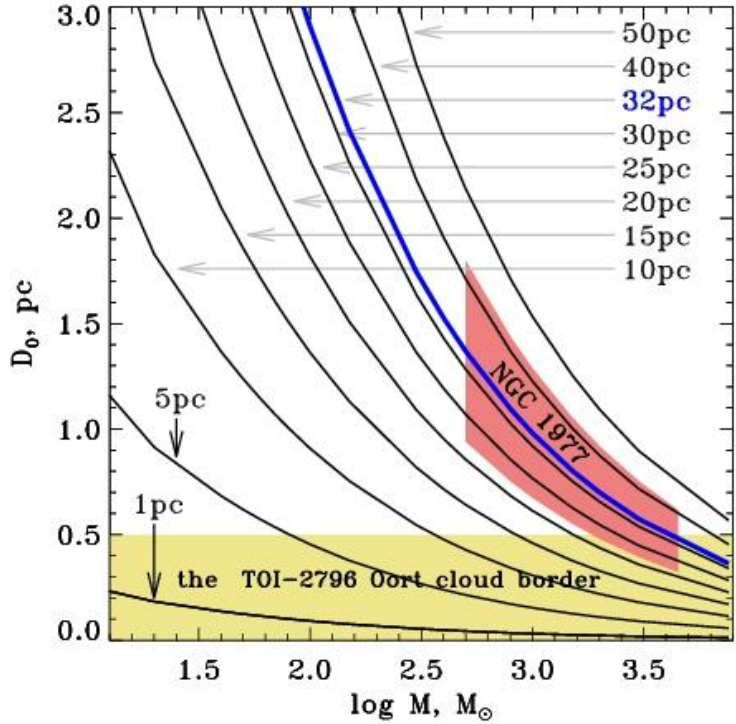

Fig. 8. Dependence of $D_0$ on $M$ according to (2) for different values of $d_{min}$ (signed in the figure). The area of the Oort cloud is shown at the bottom (the area at the bottom of the picture colored in yellow). The gravitational effect of the NGC 1977 cluster is marked by a red rectangle. According to our calculations, the center of the NGC 1977 cluster approached the host star TOI-2796 at a distance $d_{min}$ = 32.9 pc, Table 7.

It is obvious from Fig. 8 that with increasing mass of the cluster, the effect of interaction turns out to be less dependent on the mass of the cluster, provided that this mass is $\geq 1000\,M_\odot$.

The approach of NGC 1977 and TOI-2796 could have happened ~10 million years ago. The cluster age corresponds to this time also amounting to 10 million years. This may mean that the moment rapprochement corresponds to the age of accumulation. Estimating the mass of OSCs is extremely unreliable. It is connected with the difficulty of selecting stars that are member of the cluster and not completeness with faint stars. For this reason, we take the average mass estimate $\approx 1000\,M_\odot$, Lada, Lada (2003). In the region under consideration, OSCs are extremely young, located in the region of gas clouds, and represent submerged clusters. In all likelihood, they contain gas. Gas mass estimates range from 20 to ~$2000M_\odot$, Chavarria et al. (2008) and even up to ~4800 $M_\odot$ inside of one OSC, Evans et al. (2009). According to Piskunov et al. (2008) the average mass of a cluster at birth is $4.5\cdot10^3 M_\odot$. Thus, on Fig. 8 should display the mass of NGC 1977 from 500 to $4.5\cdot10^3 M_\odot$ . We consider that the mass of the cluster to be practically unchanged with the time of evolution. Numerical calculations of N bodies over the interval of ten billion years can testify to this, Sedda, Kamlah, Spurzem, Giersz et al. (2023). The cluster mass we will increase by a factor of 1.15 (Seleznev et al. 2018), assuming that 20 per cent of the cluster stars are components of binary stars.

# CONCLUSIONS

The cluster NGC 1977 was studied according to the catalog data VC2023. The age of the cluster, spatial velocity, dispersion of residual velocities, and mass were determined. A search was made for stars with exoplanets in the vicinity of 100 parsecs from the center of the cluster.

According to our calculations, the NGC 1977 cluster were closest to the host star TOI-2796 to a distance $d_{min} = 32.5$ pc. Thus, the result of our work is that the we found approach of the host star to the cluster entailed effects associated with the gravitational influence of the cluster on the nuclei of comets located in the outer parts of the Oort cloud of the TOI-2796 planetary system. The result of this influence, on the one hand, may be the appearance of comets, which in later eras could approach the star, on the other hand, some cometary nuclei could leave the planetary system and end up in interstellar space. This depends on whether the directions of the velocity vectors of the star and the cluster were co-directional or opposite.

Equatorial coordinates at the moment of approach are in Table 8.

Table 7. Distance from the Sun (R), Cartesian (X Y Z), galactic (l, b) and equatorial coordinates (α, δ) of the TOI-2796 and NGC 1977 at the moment of approach.

| Name | R, pc | X, pc | Y, pc | Z, pc | l, deg | b, deg | α, deg | δ, deg |
|------|-------|-------|-------|-------|--------|--------|--------|--------|
| TOI-2796 | 525.6 | -517.00 | -73.84 | -49.97 | 351.887 | -5.455 | 266.988 | -38.690 |
| NGC1977 | 531.75 | -525.97 | -45.10 | -63.86 | 355.099 | -6.898 | 270.572 | -36.634 |

It is important to note that the result of the approach depends on the elements of the comet's orbit: the eccentricity, inclination, and semi-major axis of the orbit. Obviously, the change in orbital velocity of the comet $\delta V_p$ in an elliptical orbit depends on its position at the moment of the approach. The place of approach in the sky is shown in Table 8 - this point can be considered as the place of appearance of interstellar comets. Thus, the result of our work is that the we found approach of the host star to the cluster entailed effects associated with the gravitational influence of the cluster on the nuclei of comets located in the outer parts of the Oort cloud of the planetary system.

# ACKNOWLEDGMENTS

This work used data from the European Space Agency (ESA) Gaia mission (https://www.cosmos.esa.int/gaia) processed by the Gaia Data Processing and Analysis Consortium (DPAC, https://www.cosmos.esa.int/web/gaia/dpac/consortium). Funding for DPAC was provided by national institutions, in particular institutions participating in the Gaia Multilateral Agreement. Mission Gaia website: https://www.cosmos.esa.int/gaia. Gaia Archive website: ttps://archives.esac.esa.int/gaia. This study used the SIMBAD database (http://cds.u-strasbg.fr) operated by CDS, Strasbourg, France. The authors are grateful to Grigory Tsurikov for useful advice.